\documentclass[11pt]{article}
\usepackage{amssymb}
\usepackage{amsmath}
\usepackage{amscd}
\usepackage{latexsym}
\catcode`@=11
\renewcommand{\section}{\@startsection{section}{1}{0pt}{\medskipamount}
{\medskipamount}{\large\bf}}
\numberwithin{equation}{section}
\catcode`@=12

\def\de{\delta}
\def\eps{\epsilon}
\def\ve{\varepsilon}

\def\h{\eta}
\def\th{\theta}

\def\l{\lambda}

\def\r{\rho}
\def\s{\sigma}
\def\p{\phi}

\def\La{\Lambda}

\newcommand{\C}{\mathbb C}
\newcommand{\R}{\mathbb R}

\newcommand{\N}{\mathbb N}
\newcommand{\Hcal}{{\cal H}}
\newcommand{\Pcal}{{\cal P}}
\newcommand{\U}{{\cal U}}

\def\e{\mbox{e}}
\def\i{\mbox{i}}
\def\N2{$N{=}2$}
\def\pa{\mbox{$\partial$}}
\def\diff{\mbox{d}}

\def\sfrac#1#2{{\textstyle\frac{#1}{#2}}}
\def\rd#1{\buildrel{_{_{\hskip 0.01in}\rightarrow}}\over{#1}}
\def\ld#1{\buildrel{_{_{\hskip 0.01in}\leftarrow}}\over{#1}}

\newcommand{\fh}{\hat{f}}
\newcommand{\gh}{\hat{g}}

\def\>{\rangle}
\def\<{\langle}
\def\+{\dagger}
\begin{document}

\begin{titlepage}
\setcounter{page}{0}
\begin{flushright}
hep-th/0109209\\
ITP--UH--24/01\\
YITP-SB-01-54\\
\end{flushright}

\vskip 2.0cm

\begin{center}

{\Large\bf  

Noncommutative 't Hooft Instantons 

}

\vspace{14mm}

{\large 
Olaf Lechtenfeld~$^{+\circ}$
\ \ and\ \ 
Alexander D. Popov~$^{\times*}$ }
\\[5mm]
{${}^+\ $ \em 
C.N. Yang Institute for Theoretical Physics \\
State University of New York \\
Stony Brook, NY 11794-3840, USA }
\\[5mm]
{${}^{\times}\ $ \em 
Institut f\"ur Theoretische Physik  \\
Universit\"at Hannover \\
Appelstra\ss{}e 2, 30167 Hannover, Germany }
\\[5mm]
{Email: lechtenf, popov@itp.uni-hannover.de}

\end{center}

\vspace{15mm}

\begin{abstract}

\noindent
We employ the twistor approach to the construction of $U(2)$ multi-instantons 
\`a la 't~Hooft~on noncommutative~$\R^4$.  The noncommutative deformation of 
the Corrigan-Fairlie-'t~Hooft-Wilczek ansatz is derived.  However, naively
substituting into it the 't~Hooft-type solution is unsatisfactory because
the resulting gauge field fails to be self-dual on a finite-dimensional 
subspace of the Fock space.  We repair this deficiency by a suitable 
Murray-von Neumann transformation after a specific projection of the gauge 
potential.  The proper noncommutative 't~Hooft multi-instanton field strength 
is given explicitly, in a singular as well as in a regular gauge.

\end{abstract}

\vfill

\textwidth 6.5truein
\hrule width 5.cm
\vskip.1in

{\small \noindent 
${}^*$
On leave from Bogoliubov Laboratory of Theoretical Physics, JINR,
Dubna, Russia\\
${}^\circ$
On sabbatical leave from Institut f\"ur Theoretische Physik, 
Universit\"at Hannover, Germany
}

\end{titlepage}

\section{Introduction and results}

\noindent
It is plausible that at very small scales space-time coordinates are to
be replaced by some noncommutative structure. In order to realize this idea
it is necessary to merge the framework of gauge field theory with the
concepts of noncommutative geometry~\cite{Connes,Connes:1998cr,Seiberg:1999vs}.

The dynamics of nonabelian gauge fields involves field configurations
not accessible by perturbation theory of which instantons are the most
prominent (in Euclidean space-time). In order to describe the nonperturbative
structure of noncommutative gauge theory, it is therefore mandatory to
construct the noncommutative deformation of instanton configurations.

The first examples of noncommutative instantons were given by Nekrasov and
Schwarz~\cite{Nekrasov:1998ss} who modified the  
Atiyah-Drinfeld-Hitchin-Manin (ADHM) construction~\cite{Atiyah:1978ri} 
to resolve the singularities of instanton moduli space (due to zero-size
instantons).
{}Furthermore, they showed that on noncommutative $\R^4$ nonsingular 
instantons exist even for the $U(1)$ gauge group. This exemplifies the
observation that noncommutativity of the coordinates eliminates singular
behavior of field configurations.
Since then, numerous papers have been devoted to this subject~[6--23],
mostly employing the modified ADHM construction on noncommutative Euclidean 
space-time.  Other related works have appeared in [24--38].

In the present paper we focus on the noncommutative generalization of
't Hooft's multi-instanton configurations for the $U(2)$ gauge group.
Nekrasov and Schwarz~\cite{Nekrasov:1998ss} proposed to keep the form of the
't Hooft solutions of the (commutative) self-duality equations but simply
impose noncommutativity on the coordinates. 
This naive noncommutative 't~Hooft configuration suffers from a problem, 
however. As was discovered by Correa et al.~\cite{Correa:2001wv} for 
the spherically-symmetric one-instanton configuration,
the Yang-Mills field fails to be self-dual everywhere.
Technically, the deficiency originates from the appearance of a source term
in the equation for the scalar field~$\phi$ in the 
Corrigan-Fairlie-'t~Hooft-Wilczek (CFtHW) ansatz.
Here, we generalize the result of~\cite{Correa:2001wv} 
to the naive noncommutative {\it multi\/}-instanton configurations 
by deriving their source terms and discuss the singular nature of the ansatz.
For comparison, we outline the ADHM derivation of the field strength 
in an Appendix.

In the commutative case, in contrast, such source terms are absent because
the singularities of $\Box\phi$ are cancelled by the zeros of~$\phi^{-1}$.
Yet, singularities are present in the gauge potential.
However, it is well known how to remove such singularities by a
singular gauge transformation, producing for example the
Belavin-Polyakov-Schwarz-Tyupkin (BPST) instanton.
Therefore, one may wonder if a noncommutative analogue exists
which removes the source terms, thus yielding a completely {\it regular\/} 
noncommutative multi-instanton whose field strength is self-dual everywhere.

In the present paper, we answer this question in the affirmative.
We easily adapt the {\it twistor approach\/} 
(which in fact underlies the ADHM scheme~\cite{Atiyah:1978ri}) 
to the noncommutative situation,
by promoting functions to operators acting on a harmonic-oscillator Fock space.
Employing the simplest Atiyah-Ward ansatz for the matrix-valued function of the
associated Riemann-Hilbert problem, we straightforwardly {\it derive\/} the 
noncommutative generalization of the CFtHW ansatz. 

The shortcoming described above is remedied by projecting the naive 
noncommutative 't~Hooft multi-instanton field strength to the source-free 
subspace of the Fock space and then applying to it a particular Murray-von 
Neumann (MvN) transformation. Such a transformation is not unitary 
but generalizes the known commutative singular gauge transformation
and removes the troublesome source term.\footnote{
For the one-instanton configuration such a transformation was considered
by Furuuchi~\cite{Furuuchi:2001vx} in the ADHM approach.}
The gauge potential producing the projected 't~Hooft multi-instanton gauge
field cannot be obtained by the standard projection since projecting does not
commute with calculating the field strength.
The projected configuration may be termed the noncommutative 't~Hooft 
instanton in a singular gauge, but its gauge potential turns out to be
given only implicitly. Nevertheless, after the MvN transformation (on the 
projected field configuration) we obtain the correct noncommutative 
't~Hooft $n$-instanton, which contains all known explicit solutions as 
special cases.

\section{Instantons from the twistor approach}

\noindent
{\bf Definitions and notation.} 
We consider the Euclidean space $\R^4$ with the metric
$\de_{\mu\nu}$, a gauge potential $A=A_{\mu}\diff x^\mu$ and the Yang-Mills 
field $F=\diff A+A\wedge A$ with components
 $F_{\mu\nu}=\pa_{\mu}A_{\nu}-\pa_{\nu}A_{\mu}+[A_{\mu},A_{\nu}]$,
where $\pa_{\mu} :=\pa /\pa x^\mu$ and $\mu,\nu,\ldots=1,2,3,4$. The field 
$A_{\mu}$ and $F_{\mu\nu}$ take values in the Lie algebra $u(2)$.

The self-dual Yang-Mills (SDYM) equations have the form:
\begin{equation}\label{sdym1}
*F\ =\ F \qquad\Longrightarrow\qquad
\sfrac{1}{2}\ve_{\mu\nu\r\s}F_{\r\s}\ =\ F_{\mu\nu}\ ,
\end{equation}
where $*$ denotes the Hodge star operator and
$\ve_{\mu\nu\r\s}$ is the completely antisymmetric tensor in 
$\R^4$, with $\ve_{1234}=1$. Solutions of~(\ref{sdym1}) having finite
Yang-Mills action are called instantons. 
Their action
\begin{equation}
S\ =\ -\frac{1}{g^2} \int\!\;\textrm{tr}\,F\wedge*F
\end{equation}
equals $8\pi^2/g^2$ times an integer which is the topological charge
\begin{equation}
Q\ =\ -\frac{1}{8\pi^2} \int\!\;\textrm{tr}\,F\wedge F \ .
\end{equation}
By `tr' we denote the trace over the $u(2)$ gauge algebra and 
by $g$ the Yang-Mills coupling constant hidden in the 
definition of the Lie algebra components of the fields $A$ and $F$.
  
If we introduce complex coordinates
\begin{equation} \label{coco}
y=x^1+\i x^2\  ,\quad z=x^3-\i x^4\ ,\quad 
\bar y=x^1-\i x^2\  ,\quad \bar z=x^3+\i x^4
\end{equation}
and put
\begin{equation}
A_y=\sfrac{1}{2}(A_1-\i A_2)\ ,\quad 
A_z=\sfrac{1}{2}(A_3+\i A_4)\ ,\quad 
A_{\bar y}=\sfrac{1}{2}(A_1+\i A_2)\ ,\quad
A_{\bar z}=\sfrac{1}{2}(A_3-\i A_4)\ ,
\end{equation}
then the SDYM equations (\ref{sdym1}) will read
\begin{equation}\label{sdym2} 
[D_y, D_z]=0\quad ,\qquad [D_{\bar y},D_{\bar z}]=0\quad ,\qquad 
[D_y, D_{\bar y}]+[D_z, D_{\bar z}]=0\ , 
\end{equation}
where $D_\mu :=\pa_\mu +A_\mu$.
These equations can be obtained as the compatibility condition 
of the following linear system of equations:
\begin{equation}\label{ls}
(D_{\bar y} - \l D_z )\,\psi (x,\l )\ =\ 0\qquad\mbox{and}\qquad
(D_{\bar z} + \l D_y )\,\psi (x,\l )\ =\ 0\ ,
\end{equation}
where the $2\times 2$ matrix $\psi$
depends on $(y, \bar y, z, \bar z, \l )$ but not on~$\bar\l$. 
The `spectral parameter'
$\l$ lies in the extended complex plane $\C P^1=\C\cup\{\infty\}$.

\noindent
{\bf Twistors and transition functions.} 
In fact, the function $\psi$ in (\ref{ls}) is defined 
on the {\it twistor space\/} $\Pcal =\R^4\times\C P^1$ for 
the space $\R^4$~\cite{Atiyah:1977pw,Atiyah:1978wi}. The sphere 
$S^2$, considered as the complex projective
line~$\C P^1$, can be covered by two coordinate patches $U_+$ and $U_-$ with
\begin{equation}
\C P^1=U_+\cup  U_-\quad ,\qquad 
U_+=\C P^1\setminus\{\infty\}\quad ,\qquad
U_-=\C P^1\setminus\{0\}\ ,
\end{equation}
and coordinates $\l$ and $\tilde\l$ on $U_+$ and $U_-$, respectively.
Therefore, also $\Pcal$ can be covered by two coordinate patches,
\begin{equation}
\Pcal =\U_+\cup\U_-\quad ,\qquad 
\U_+=\R^4\times U_+\quad ,\qquad 
\U_-=\R^4\times U_-\ ,
\end{equation}
with complex coordinates
\begin{equation}
w_1=y-\l\bar z\ ,\ w_2=z+\l \bar y\ ,\ w_3=\l\qquad\mbox{and}\qquad
\tilde w_1=\tilde \l y-\bar z\ ,\ \tilde w_2=\tilde \l z+\bar y\ ,\ 
\tilde w_3=\tilde \l
\end{equation}
on $\U_+$ and $\U_-$, respectively. On the intersection
$\U_+\cap\U_-\simeq\R^4\times\C^*$ these coordinates are related by
\begin{equation}
\tilde w_1=\frac{w_1}{w_3}\quad ,\qquad \tilde w_2=\frac{w_2}{w_3}
\qquad\mbox{and}\qquad \tilde w_3=\frac{1}{w_3}\ .
\end{equation}
On the open set ~$\U_+\cap\U_-$ one may use any of them, and we will 
use $w_1$, $w_2$ and $w_3{=}\l$.

There exist two matrix-valued solutions $\psi_+(x,\l )$ and 
$\psi_-(x,\l )$ of~(\ref{ls}) which are defined on 
$\U_+$ and $\U_-$, respectively. Finding them, one can introduce the 
matrix-valued function
\begin{equation}\label{triv}
{}f_{+-}\ :=\ \psi_+^{-1}\,\psi_-
\end{equation}
defined on the open set $\U_+\cap\U_-\subset \Pcal$. From~(\ref{ls}) it 
follows that $f_{+-}$ depends on the complex coordinates 
$w_1$, $w_2$ and $\l$ holomorphically, 
\begin{equation}
(\pa_{\bar y}-\l\pa_z)f_{+-}=0 \quad\textrm{and}\quad
(\pa_{\bar z}+\l\pa_y)f_{+-}=0 \qquad \Longrightarrow \qquad
f_{+-}=f_{+-}(w_1,w_2,\l)\ .
\end{equation}
Any such function defines a holomorphic bundle over $\Pcal$. Namely, $f_{+-}$ 
can be identified with a transition function in a holomorphic bundle over 
$\Pcal$, and a pair of functions $\psi_{\pm}$
defines a smooth trivialization of this bundle.

\noindent
{\bf Gauge equivalence and reality conditions.} 
It is easy to see that gauge transformations
\begin{equation}
A_\mu\quad \mapsto\quad A_\mu^g\ =\ g^{-1}A_\mu\, g +g^{-1}\pa_\mu\, g
\end{equation}
are induced by the transformations
\begin{equation}
\psi_+\quad \mapsto\quad \psi_+^g\ =\ g^{-1}\psi_+\qquad\textrm{and}\qquad
\psi_-\quad \mapsto\quad \psi_-^g\ =\ g^{-1}\psi_-\ ,
\end{equation}
where $g{=}g(x)$ is an arbitrary $U(2)$-valued function on~$\R^4$.
The transition function $f_{+-}=\psi_+^{-1}\psi_-$ is invariant 
under these transformations. On the other hand, 
the gauge potential $A$ is inert under the transformations
\begin{equation} \label{inert}
\psi_+\quad\mapsto\quad\psi_+\,h_+^{-1}\qquad\textrm{and}\qquad
\psi_-\quad\mapsto\quad\psi_-\,h_-^{-1}\ ,
\end{equation}
where $h_+{=}h_+(w_1,w_2,\l)$ and 
$h_-{=}h_-(\tilde w_1,\tilde w_2,\tilde \l)$ 
are arbitrary matrix-valued 
holomorphic functions on $\U_+$ and $\U_-$, respectively.

The reality of the gauge fields is 
an important issue~\cite{Atiyah:1977pw,Atiyah:1978wi}.
The antihermiticity conditions 
$A_\mu^\+ = -A_\mu$ for components of the gauge potential imply
the following `reality' conditions for the matrices $\psi_{\pm}$ and $f_{+-}$:
\begin{equation}\label{real} 
\psi_+^\+(x, - {\bar\l}^{-1})\ =\ \psi_-^{-1}(x,\l) \qquad\textrm{and}\qquad
f_{+-}^\+ (x, -{\bar\l}^{-1})\ =\ f_{+-} (x,\l)\ .
\end{equation}

\noindent
{\bf Splitting of transition functions.} 
Consider now the inverse 
situation. Let us have a holomorphic matrix-valued function $f_{+-}$ 
on the open subset $\U_+\cap\U_-$ of the twistor space $\Pcal$. 
Suppose we are able to split $f_{+-}$, i.e. for each fixed 
$x\in\R^4$ find matrix-valued functions $\psi_{\pm}(x,\l )$ such 
that $f_{+-}=\psi_+^{-1}\psi_-$ on $U_+\cap U_-$ and the functions $\psi_+$ and 
$\psi_-$ can be extended continuously to functions regular on $U_+$ and 
$U_-$, respectively. From the holomorphicity of $f_{+-}$ it then follows 
that
\begin{equation}\label{hol}
\psi_{+}(\pa_{\bar y}-\l\pa_z)\psi_{+}^{-1} =
\psi_{-} (\pa_{\bar y}-\l\pa_z)\psi_-^{-1} \quad\ \textrm{and}\quad\
\psi_{+}(\pa_{\bar z}+\l\pa_y)\psi_{+}^{-1} =
\psi_{-} (\pa_{\bar z}+\l\pa_y)\psi_-^{-1}\ .
\end{equation}

Recall that the matrix-valued functions $\psi_+$ and $\psi_-$ are regular 
on their respective domains, so that we may expand them
into power series in $\l$ and $\l^{-1}$, respectively,
$\psi_{\pm}=\sum_{n\ge 0}\l^{\pm n}\psi_{\pm,n}(x)$.
Upon substituting into~(\ref{hol}) one easily sees that both sides 
of~(\ref{hol}) must be linear in $\l$, and one can introduce $A_\mu$ by
\begin{equation}\label{A1}
A_{\bar y}-\l A_z\ = \psi_{\pm}(\pa_{\bar y}-\l\pa_z)\psi_{\pm}^{-1}
\qquad\textrm{and}\qquad
A_{\bar z}+\l A_y\ = \psi_{\pm}(\pa_{\bar z}+\l\pa_y)\psi_{\pm}^{-1} \ .
\end{equation}
Hence, the gauge field components may be calculated from
\begin{equation}\label{A2}
A_{\bar y}\ =\ \psi_{+}\pa_{\bar y}\psi_{+}^{-1}|_{\l =0}\ =\
-A_y^\+ \qquad\textrm{and}\qquad
A_{\bar z}\ =\ \psi_{+}\pa_{\bar z}\psi_{+}^{-1}|_{\l =0}\ =\
-A_z^\+ \ .
\end{equation}
By construction, the components $\{A_\mu\}$ of the gauge potential $A$ 
defined by (\ref{A1}) or (\ref{A2}) satisfy the SDYM equations.
{}For more detailed discussion of {\it local\/} solutions, their 
infinite-dimensional moduli space and references see 
e.g.~\cite{Crane:1987im,Popov:1999pc}.

{}For a fixed point $x\in\R^4$, the task to split a matrix-valued holomorphic 
function $f_{+-}\equiv f_{+-}(y{-}\l\bar z, z{+}\l\bar y, \l )$
defines a {\it parametric Riemann-Hilbert problem\/} on $\C P^1$. 
The explicit general solution of this Riemann-Hilbert problem is not known. 
{}For a large class of special cases, however, the splitting can be achieved. 
In this paper we shall make explicit use of a particular example,
the so-called Atiyah-Ward ansatz~\cite{Atiyah:1977pw}, to be presented
momentarily in the noncommutative context.

\section{The Atiyah-Ward ansatz}

\noindent
{\bf Noncommutative Yang-Mills theory.}
The noncommutative deformation of (classical) field theory is most easily
effected by extending the function product in field space to the star product
\begin{equation}
(f \star g)(x)\ =\ f(x)\,\exp\,\bigl\{ \frac{\i}{2}
{\ld{\partial}}_\mu \,\theta^{\mu\nu}\, {\rd{\partial}}_\nu \bigr\}\,g(x) \ ,
\end{equation}
with a constant antisymmetric tensor~$\th^{\mu\nu}$.
In this work, we restrict ourselves to the case of a self-dual ($\eps =1$) or
a anti-self-dual ($\eps =-1$) tensor $\th^{\mu\nu}$ and 
choose coordinates such that
\begin{equation} \label{asdth}
\th^{12}\ =\ -\th^{21}\ =\ \eps\th^{34}\ =\ -\eps\th^{43}\ =\ \th\ >0 \ .
\end{equation}
In star-product formulation, the SDYM equations~(\ref{sdym1}) are formally 
unchanged, but the components of the noncommutative field strength now read
\begin{equation}
{}F_{\mu\nu}\ =\ \pa_\mu A_\nu-\pa_\nu A_\mu+A_\mu\star A_\nu-A_\nu\star A_\mu\ .
\end{equation}

The nonlocality of the star product renders explicit computations cumbersome.
We therefore take advantage of the Moyal-Weyl correspondence and
pass to the operator formalism,
which trades the star product for operator-valued coordinates
$\hat{x}^\mu$ satisfying $[\hat{x}^\mu,\hat{x}^\nu]=\i\theta^{\mu\nu}$.
This defines the noncommutative Euclidean space $\R_\th^4$.
In complex coordinates~(\ref{coco}) our choice~(\ref{asdth}) implies
\begin{equation} \label{ncco}
[\,\hat{y}\,,\,\hat{\bar{y}}\,]\ =\ 2\th \quad,\qquad
[\,\hat{z}\,,\,\hat{\bar{z}}\,]\ =\ -2\eps\th \quad,\qquad
\textrm{and other commutators}\ =\ 0 \ .
\end{equation}
Clearly, coordinate derivatives are now inner derivations of this algebra, i.e.
\begin{equation}\label{code}
\hat{\pa}_y \fh\ =\ \sfrac{-1}{2\th}\,[\,\hat{\bar y}\,,\,\fh\,] 
\qquad\textrm{and}\qquad
\hat{\pa}_{\bar y} \fh\ =\ \sfrac{1}{2\th}\,[\,\hat{y}\,,\,\fh\,] 
\end{equation}
for any function~$\fh$ of $(\hat y,\hat{\bar y},\hat z,\hat{\bar z})$.
Analogous formulae hold for $\hat{\pa}_z$ and $\hat{\pa}_{\bar z}$.

The obvious representation space for the Heisenberg algebra~(\ref{ncco})
is the two-oscillator Fock space~$\Hcal$ spanned by 
$\{ |n_1,n_2\>\ \textrm{with}\ n_1,n_2=0,1,2,\ldots\}$.
In $\Hcal$ one can introduce an integer ordering of states e.g. 
as follows~\cite{Rangamani:2001cn}:
\begin{equation} \label{fock}
|k\>\ =\ |n_1,n_2\> \ =\
\frac{\hat{\bar y}^{n_1}\,\hat{\bar z_\eps}^{n_2}\,|0,0\>}
{\sqrt{n_1!n_2!(2\th)^{n_1+n_2}}}
\qquad\textrm{with}\qquad
\hat{z}_\eps\ :=\ 
\sfrac{1-\eps}{2}\,\hat{z}\ +\ \sfrac{1+\eps}{2}\,\hat{\bar{z}}
\end{equation}
and $k=n_1+\sfrac{1}{2}(n_1+n_2)(n_1+n_2+1)$.
Thus, coordinates as well as fields are to be regarded as 
operators in~$\Hcal$. The Moyal-Weyl map yields the operator equivalent 
of star multiplication and integration,
\begin{equation} \label{trace}
f\star g\ \longmapsto\ \fh\,\gh \qquad\textrm{and}\qquad
\int\! \diff^4{x}\,f\ =\
(2\pi \theta)^2 \,\mbox{Tr}_\Hcal\, \fh\ ,
\end{equation} 
respectively, where `$\mbox{Tr}_\Hcal$' signifies the trace 
over the Fock space~$\Hcal$.

In the operator formulation,
the noncommutative generalization of the SDYM equations~(\ref{sdym2})
again retains their familiar form,
\begin{equation}\label{sdym3}
\hat{F}_{yz}=0\quad ,\qquad \hat{F}_{\bar y\bar z}=0\quad ,\qquad  
\hat{F}_{ y\bar y}+\hat{F}_{z\bar z}=0\ .
\end{equation}
The operator-valued field-strength components $\hat{F}_{\mu\nu}$, however, 
now relate to the noncommutative gauge-potential components $\hat{A}_\mu$ 
with the help of~(\ref{code}), as e.g. in
\begin{equation}
2\th\,\hat{F}_{yz}\ =\ 
[-\hat{\bar{y}}+\th\hat{A}_y,\hat{A}_z] -
[\eps\hat{\bar{z}}+\th\hat{A}_z,\hat{A}_y] \ .
\end{equation}
{}For the rest of the paper we shall work in the operator formalism and 
drop the hats over the operators in order to avoid cluttering the notation.

\noindent
{\bf Noncommutative Atiyah-Ward ansatz.} 
Commutative instantons can be obtained by the famous 
ADHM construction~\cite{Atiyah:1978ri},
which was derived from the twistor approach.
Almost all works on noncommutative instantons are based on the
modified ADHM construction~\cite{Nekrasov:1998ss}.
At the same time,
it is known that the modified ADHM construction can be interpreted 
in terms of a noncommutative version of the twistor 
transform~\cite{Kapustin:2001ek,Takasaki:2001vs}.
Therefore it is reasonable to expect that an approach based on 
the splitting of transition functions in a 
holomorphic bundle over a noncommutative twistor 
space~\cite{Kapustin:2001ek,Takasaki:2001vs,Hannabuss:2001xj} will 
work as well. Here we show that this is indeed the case 
for the simplest Atiyah-Ward ansatz for the transition functions. 

In the commutative situation the infinite hierarchy of 
Atiyah-Ward ans\"atze reads
\begin{equation}\label{AW}
f^{(k)}_{+-}\ =\ \begin{pmatrix} \l^k & 0 \\ \r & \l^{-k} \end{pmatrix}\ ,
\end{equation}
where $k=1,2,\ldots$ and $\r$ denotes a holomorphic {\it function\/}
on~$\U_+\cap\U_-\subset \Pcal$.\footnote{
A reminder on the commutative situation:
{}For $k{=}1$ the ansatz (\ref{AW}) leads to a parametrization of
self-dual gauge fields in terms of a scalar field
$\phi = \textrm{res}_{\l =0} (\l^{-1}\r )$ satisfying the wave equation,
while for $k{\ge}2$ it produces solutions more general than
the 't~Hooft $n$-instanton
configurations~\cite{Atiyah:1977pw,Corrigan:1978ma,Prasad:1980yy,Ward:1981kj}.
Note that the matrices $f^{(k)}_{+-}$ do not
satisfy the reality condition~(\ref{real}).
}
We are confident that 
the ans\"atze~(\ref{AW}) allow one to construct solutions of the
noncommutative SDYM equations~(\ref{sdym3}) simply by promoting $\r$ to an 
{\it operator\/} acting in the Fock space $\Hcal$.  

Specializing to $k=1$, we introduce the matrix
\begin{equation}\label{AW1}
f_{+-}\ :=\ 
\left(\begin{smallmatrix} 0 & +1 \\[4pt] -1 & 0 \end{smallmatrix}\right)\
f^{(1)}_{+-}\ =\ 
\begin{pmatrix} \r & {\l}^{-1} \\ -\l & 0 \end{pmatrix}\ .
\end{equation}
Because $f_{+-}$ is related to $f^{(1)}_{+-}$ 
by a transformation~(\ref{inert}), 
with $h_+=(\begin{smallmatrix} 0&1\\ \!{-}1&0 \end{smallmatrix})$
and  $h_-=(\begin{smallmatrix} 1&0\\0&1 \end{smallmatrix})$,
it leads to the same gauge field configuration.  Yet, 
$f_{+-}$ has the advantage of satisfying the reality condition~(\ref{real}). 
Here $\r$ is a holomorphic `real' operator-valued function, i.e.
\begin{equation}\label{rho}
(\pa_{\bar y}-\l\pa_z)\r\ =\ (\pa_{\bar z}+\l\pa_y)\r\ =\ 0 
\qquad\textrm{and}\qquad
\r^\+(x,-\bar\l^{-1})\ =\ \r(x,\l)\ .
\end{equation}
It is useful to expand $\r$ in a Laurent series in $\l$,
\begin{equation}\label{sum}
\r\ = \sum\limits^\infty_{m=-\infty}\r_m\,\l^m\ =\ \r_-+\phi +\r_+\ ,
\end{equation}
where
\begin{equation}
\r_+:=\sum\limits_{m >0}\r_m\,\l^m \quad ,\qquad 
\r_-:=\sum\limits_{m <0}\r_m\,\l^m \qquad\textrm{and}\qquad 
\phi:=\r_0\ .
\end{equation}
The reality condition~(\ref{real}) then becomes
\begin{equation}
\phi^\+(x)\ =\ \phi (x) \qquad\mbox{and}\qquad
\r_+^\+(x, -{\bar\l}^{-1})\ =\ \r_-(x,\l ) \ .
\end{equation}

It is not difficult to see that $f_{+-}$ can be split as
$f_{+-}=\psi^{-1}_+\psi_-$, where
\begin{equation}
\psi_-\ =\ \frac{1}{\sqrt\phi}
\begin{pmatrix} \phi+\r_- & {\l}^{-1} \\ \l\r_- & 1 \end{pmatrix}
\qquad\textrm{and}\qquad
\psi^{-1}_+\ =\ 
\begin{pmatrix} \phi+\r_+ & -{\l}^{-1}\r_+ \\ -\l & 1 \end{pmatrix} 
\frac{1}{\sqrt\phi}\ .
\end{equation}
Consequently we have
\begin{equation}
\psi^{-1}_-\ =\ 
\begin{pmatrix} 1 & -{\l}^{-1} \\ -\l\r_- & \phi+\r_- \end{pmatrix}
\frac{1}{\sqrt\phi}
\qquad\textrm{and}\qquad
\psi_+\ =\ \frac{1}{\sqrt\phi}
\begin{pmatrix} 1 & {\l}^{-1}\r_+ \\ \l & \phi+\r_+ \end{pmatrix}\ ,
\end{equation}
which satisfy the reality condition~(\ref{real}).
Thus, the first Atiyah-Ward ansatz is easily generalized to the 
noncommutative case.

\noindent
{\bf Parametrization of the gauge potential.} 
{}From the operator version of formulae (\ref{A1})--(\ref{A2}) and 
recursion relations 
\begin{equation}\label{rel}
\pa_{\bar y}\,\r_{m+1}\ =\ \pa_z\r_m \qquad\mbox{and}\qquad
\pa_{\bar z}\,\r_{m+1}\ =\ -\pa_y\r_m
\end{equation}
implied by~(\ref{rho}) for the Laurent coefficients in~(\ref{sum})
we get
\begin{align}
A_{\bar y}\ &=\ \begin{pmatrix}
\phi^{-\frac{1}{2}}\pa_{\bar y}\phi^{\frac{1}{2}} &
-\phi^{-\frac{1}{2}} (\pa_{z}\phi )\phi^{-\frac{1}{2}} \\ 
0 &
\phi^{\frac{1}{2}}\pa_{\bar y}\phi^{-\frac{1}{2}}
\end{pmatrix} \quad ,\qquad
A_{\bar z}\ =\ \begin{pmatrix}
\phi^{-\frac{1}{2}}\pa_{\bar z}\phi^{\frac{1}{2}} &
\phi^{-\frac{1}{2}} (\pa_{y}\phi )\phi^{-\frac{1}{2}} \\ 
0 &
\phi^{\frac{1}{2}}\pa_{\bar z}\phi^{-\frac{1}{2}}
\end{pmatrix} \quad, \nonumber \\[8pt] \label{ncgen}
A_{y}\ &=\ \begin{pmatrix}
\phi^{\frac{1}{2}}\pa_{y}\phi^{-\frac{1}{2}} &
0 \\
\phi^{-\frac{1}{2}}(\pa_{\bar z}\phi )\phi^{-\frac{1}{2}} &
\phi^{-\frac{1}{2}}\pa_{y}\phi^{\frac{1}{2}}
\end{pmatrix} \quad ,\qquad
A_{z}\ =\ \begin{pmatrix}
\phi^{\frac{1}{2}}\pa_{z}\phi^{-\frac{1}{2}} &
0 \\ 
-\phi^{-\frac{1}{2}}(\pa_{\bar y}\phi )\phi^{-\frac{1}{2}} &
\phi^{-\frac{1}{2}}\pa_{z}\phi^{\frac{1}{2}}
\end{pmatrix} \quad.
\end{align}
Rewriting these expressions in real coordinates $x^\mu$, we obtain
the noncommutative generalization of the CFtHW ansatz,
\begin{equation}\label{Amu}
A_\mu\ =\ \bar\eta^a_{\mu\nu}\frac{\s_a}{2\i}
\left(\p^{\frac12}\pa_\nu\p^{-\frac12}-\p^{-\frac12}\pa_\nu\p^{\frac12}\right)
+ \frac{{\mathbf 1}_2}{2}
\left(\p^{-\frac12}\pa_\mu\p^{\frac12}+\p^{\frac12}\pa_\mu\p^{-\frac12}\right)
\ ,
\end{equation}
where
\begin{equation}
\bar\eta^a_{\mu\nu}\ =\ \begin{cases}
\eps^a_{bc} & \textrm{for} \quad \mu =b\,,\ \nu =c \\
-\de^a_\mu  & \textrm{for} \quad \nu =4 \\
\de^a_\nu   & \textrm{for} \quad \mu =4 \end{cases}
\end{equation}
is the anti-self-dual 't~Hooft tensor~\cite{Prasad:1980yy}, and 
\begin{equation}
\s_1=\begin{pmatrix}0&1\\1&0\end{pmatrix}\quad ,\qquad
\s_2=\begin{pmatrix}0&-\i\\\i&0\end{pmatrix}\quad ,\qquad
\s_3=\begin{pmatrix}1&0\\0&-1\end{pmatrix}
\end{equation}
are the Pauli matrices.

\noindent
{\bf Reduced SDYM equation.} 
Calculating the Yang-Mills curvature for the 
noncommutative CFtHW ansatz~({\ref{Amu}) results in
\begin{equation}\label{csdym}
{}F_{\bar y\bar z}=\begin{pmatrix}0&X\\ 0&0\end{pmatrix}\quad,\qquad
{}F_{yz}          =\begin{pmatrix}0&0\\-X&0\end{pmatrix}\quad,\qquad
{}F_{ y\bar y}+F_{z\bar z}=\begin{pmatrix}X&0\\0&-X\end{pmatrix}\ ,
\end{equation}
with
\begin{equation}
X\ =\ \phi^{-\frac12} 
(\pa_y\pa_{\bar y}\phi + \pa_z\pa_{\bar z}\phi )\,\phi^{-\frac12}\ .
\end{equation}
Hence, for the ansatz (\ref{Amu}) the noncommutative SDYM 
equations~(\ref{sdym3}) are reduced to 
\begin{equation}\label{redop}
\phi^{-\frac12}
(\pa_y\pa_{\bar y}\phi + \pa_z\pa_{\bar z}\phi )\,\phi^{-\frac12}\ =\ 0\ .
\end{equation}

It is natural to assume~\cite{Nekrasov:1998ss} that a solution $\phi_n$ of 
this equation looks exactly like the standard 't~Hooft solution 
\begin{equation}\label{phin}
\phi_n\ =\ 1 +\,\sum^n_{i=1}\frac{\Lambda^2_i}{(x^\mu-b_i^\mu)(x^\mu-b_i^\mu)}\
=:\ 1+\,\sum_{i=1}^n\frac{\Lambda^2_i}{r^2_i}\ ,
\end{equation}
where now the $x^\mu$ are operators and
$$
\frac{1}{ x^\mu x^\mu}\ :=\ \frac{1}{2\theta}\sum_{{n_1},{n_2}\ge 0}
\frac{1}{n_1{+}n_2{+}1}\,|n_1,n_2\>\<n_1,n_2|\ ,
$$
$$
r_i^2\ =\ \de_{\mu\nu}(x^\mu -b_i^\mu )(x^\nu -b_i^\nu )\
=\ \bar y_iy_i+z_{i\eps}\bar z_{i\eps}\
=\ \bar y_iy_i+\bar z_{i\eps}z_{i\eps}+ 2\th\ ,
\phantom{XXX}
$$
\begin{equation}
y_i:=y-b_i^y\ ,\quad z_{i\eps}:=z_\eps-b^z_i\ ,\quad 
b_i:=(b_i^\mu)=(b_i^y,b_i^z)\ , \quad 
\bar b_i:=(\bar b_i^\mu)=(\bar b_i^y, \bar b_i^z)\ .
\end{equation}
The real parameters $b_i^\mu$ and $\Lambda_i$ denote the position coordinates 
and the scale of the $i$th instanton.
In particular, as a candidate for the one-instanton solution we have 
(putting $b_1^\mu{=}0$)
\begin{equation}\label{phi1}
\phi_1\ =\ 1+\frac{\Lambda^2}{r^2} \ ,
\end{equation}
where 
\begin{equation} \label{ncr}
r^2\ =\ \de_{\mu\nu}x^\mu x^\nu\ =\ 
\sfrac{1}{2}(\bar yy+y\bar y+\bar zz+z\bar z)\ =\ 
\bar yy + z_\eps\bar z_\eps\ =\
\bar yy + \bar z_\eps z_\eps + 2\th\ .
\end{equation}

\noindent
{\bf Sources.}
When checking the self-duality of the purported one-instanton configuration
one discovers a subtlety:
The substitution of~(\ref{phi1}) into the reduced SDYM equation~(\ref{redop}) 
and results~\cite{Gross:2000wc} on Green's functions reveal 
that the deviation from self-duality, 
\begin{equation}\label{ex}
X_1\ :=\ \phi^{-\frac12}_1
(\pa_y\pa_{\bar y}\phi_1 + \pa_z\pa_{\bar z}\phi_1 )\,\phi^{-\frac12}_1\
=\ -\frac{\Lambda^2}{2\th (\Lambda^2 +2\th )}\,P_0
\qquad \mbox{with}\quad P_0:=|0,0\>\<0,0| \ ,
\end{equation}
is not zero if $\Lambda^2{\neq}0$.  
This has led the authors of~\cite{Correa:2001wv} 
to the conclusion that this configuration is {\it not\/} self-dual. 
This is not the whole story - one can 
show that all $\phi_n$ in~(\ref{phin}) fail to satisfy~(\ref{redop}).
Namely, by differentiating we obtain
\begin{equation}\label{ddphin}
\pa_y\pa_{\bar y}\phi_n+\pa_z\pa_{\bar z}\phi_n\ =\ -\frac{1}{4\th^2}
\sum_{i=1}^n \Lambda^2_i\, |b_i\>\<b_i|\ ,
\end{equation}
where
\begin{equation} \label{bstates}
|b_i\>\ :=\ |b_i^y, b_i^z\>\ =\ \e^{-\frac{1}{2}|b_i|^2}\, 
\e^{b_i^y\bar y/\sqrt{2\th}}\,\e^{b_i^z \bar z_\eps/\sqrt{2\th}}\, |0,0\>
\end{equation}
denotes the `shifted ground state' centered at $(b_i^\mu)$ which is
constructed as a coherent state for the Heisenberg-Weyl group
generated by the algebra~(\ref{ncco}).  The vectors $|b_i\>,\ i=1,\ldots,n$, 
span an $n$-dimensional subspace of~$\Hcal$. Since they are not orthonormalized
it is useful to also introduce an orthonormal basis $\{|h_1\>,\ldots,|h_n\>\}$ 
of this subspace through~\cite{Rocek}
\begin{equation}
\bigl(\,|h_1\>,\,|h_2\>,\ldots,\,|h_n\>\,\bigr)\ :=\ T\,(T^\+ T)^{-\frac12}
\qquad\textrm{with}\qquad
T\ :=\ \bigl(\,|b_1\>,\,|b_2\>,\ldots,\,|b_n\>\,\bigr) \ .
\end{equation}
This basis can be extended to an orthonormal basis of the whole Fock space
$\Hcal$ by simply adjoining further vectors $|h_{n+1}\>,\,|h_{n+2}\>,\ldots\ $.
{}From the explicit form of the function~$\phi_n$ it follows that the
matrix elements $\<h_i|\phi^{-\frac{1}{2}}_n|h_j\>$ 
are nonzero only for $i,j\le n$.
Therefore, the operator $\phi^{-\frac{1}{2}}_n$ preserves 
the subspace of~$\Hcal$ spanned by the vectors $|b_i\>,\ i=1,\ldots,n$, i.e.
\begin{equation}
\phi^{-\frac{1}{2}}_n|b_i\>\ =\ \sum_{j=1}^n M_{ij}\,|b_{j}\>
\qquad\textrm{and}\qquad
\<b_i|\,\phi^{-\frac{1}{2}}_n\ =\ \sum_{k=1}^n\<b_k|\,\overline{M}_{ki}
\end{equation}
where $M_{ij} = M_{ij}(\th,\Lambda^2_1,\ldots,\Lambda^2_n)$ are some constants.
Using this fact we arrive at
\begin{equation}\label{x}
X_n\ :=\ \phi_n^{-\frac12}(\pa_y\pa_{\bar y}\phi_n+\pa_z\pa_{\bar z}\phi_n)\,
\phi_n^{-\frac12}\ =\ -\frac{1}{4\th^2}\sum_{i,j,k=1}^n |b_j\>\,
\overline{M}_{ki}\,\Lambda^2_i\,M_{ij}\,\<b_k|\ ,
\end{equation}
which is not zero if at least one $\Lambda^2_i$ does not vanish.
The Appendix derives the same expression via the ADHM approach.

\vfill\eject

\section{Noncommutative instantons}

\noindent
{\bf Projected field configurations.}
The twistor approach has given us a systematic way of producing 
a noncommutative generalization~(\ref{Amu}) of the CFtHW ansatz.
However, it has been shown above that the noncommutative 't Hooft
type ansatz~(\ref{phin}) does not solve the SDYM equations since
$X_n\not=0$ produce sources in the r.h.s. of~(\ref{csdym}).
These sources are localized on a finite-dimensional subspace of
the Fock space.

Let us look at the situation in more detail.
Recall that  in the noncommutative case the
components $A_\mu$ and $F_{\mu\nu}$ are {\it operators\/} acting 
(on the left) in the space $\Hcal\otimes\C^2 = \Hcal\oplus\Hcal$ 
which carries a fundamental representation of the group $U(2)$. 
It is easy to see that for the ansatz~(\ref{phi1}) each
term of the operators $A_\mu$ in~(\ref{Amu}) 
annihilates the state $|0,0\>\otimes\C^2$ or $\C^2\otimes\<0,0|$ 
(when acting on the right), thus $\<0,0|A_\mu |0,0\>=0$. 
This shows that the $A_\mu$ are well defined only on the subspace
$(1{-}P_0)\Hcal\otimes \C^2$ of the Fock space.
Moreover, from~(\ref{ex}) we see that the gauge fields are self-dual
only on the same subspace since $X_1(1{-}P_0)=0$. 
Analogously, one can easily see that $X_n$ given by~(\ref{x})
is annihilated by the projector
\begin{equation}
1-P_{n-1}\ :=\ 1-\sum_{i,j=1}^n|b_i\>\frac{1}{\<b_i|b_j\>}\<b_j|
\ =\ 1-\sum_{i=1}^n\,|h_i\>\<h_i|
\end{equation}
onto the orthogonal complement of the subspace spanned by $\{|b_i\>, 
i=1,\ldots,n\}$ but not outside it. Therefore, the solutions based 
on~(\ref{phin}) are self-dual on the reduced Fock 
space~$(1{-}P_{n-1})\Hcal\otimes\C^2$ (projective Fock module).
The same phenomenon occurs in the modified ADHM construction 
of the noncommutative one- and two-instanton solutions and was 
discussed intensively by Ho~\cite{Ho:2000ea} and especially by Furuuchi
\cite{Furuuchi:2000kv,Furuuchi:2001vx,Furuuchi:2001dx,Furuuchi:2000vc}. 
The ADHM construction of noncommutative 't~Hooft $n$-instantons including 
the appearance of a source term for any~$n$ is described in the Appendix.
Note that the deletion of a subspace generated by $n$~states 
$|b_i\>,\ i=1,\ldots,n,$ from the Fock space corresponds to the exclusion 
of $n$~points $b_i,\ i=1,\ldots,n,$ (in which the gauge potential 
is singular) from the commutative space~$\R^4$.
Hence, the sources in the r.h.s. of~(\ref{csdym}), being localized on the
subspace $P_{n-1}\Hcal\otimes\C^2$,
imply that the ansatz~(\ref{phin}) produces a {\it singular\/} solution.

Quite generally, consider any finite-rank projector~$P$.
If the deviation from self-duality is localized on the subspace 
$P\Hcal\otimes\C^2$ of the Fock space then obviously 
we can produce a self-dual field strength $F^p_{\mu\nu}$
by projecting the gauge field onto the reduced
Fock space $(1{-}P)\Hcal\otimes\C^2$,  
\begin{equation}\label{FP1}
F_{\mu\nu}\quad\mapsto\quad F^p_{\mu\nu}\ =\ (1{-}P)\,F_{\mu\nu}\,(1{-}P)\ .
\end{equation}
Unfortunately, the corresponding gauge potential~$A^p_\mu$ generating this 
$F^p_{\mu\nu}$ {\it cannot\/} be obtained by naively projecting $A_{\mu}$, 
since projecting does not commute with calculating the field strength:
$$
\begin{CD}
A @>{\delta}>> F \\
@V{?}VV @VV{1{-}P}V \\
A^p @>{\delta_p}>> F^p
\end{CD}
$$
where $\ {\downarrow}\,{\scriptstyle1{-}P}\ $ denotes the standard projection
onto $(1{-}P)\Hcal\otimes\C^2$ and $\delta$ and $\delta_p$ compute 
the curvature in the full and the reduced Fock space, respectively. 
More explicitly, it reads
\begin{align}
F_{\mu\nu}\ &=\ \pa_{[\mu} A_{\nu]} + [A_\mu\,,A_\nu] \ , \nonumber\\[8pt]
F^p_{\mu\nu}\ &=\ (1{-}P)\,\pa^{\phantom{p}}_{[\mu} A^p_{\nu]}\,(1{-}P) +
[A^p_\mu\,,A^p_\nu] + (1{-}P)\,\pa_{[\mu}P\,\pa_{\nu]}P \ . \label{FP2}
\end{align}

Although $A^p$ is a projected gauge potential, i.e. it fulfils
\begin{equation}
A^p_\mu\ =\ (1{-}P)\,A^p_\mu\,(1{-}P) \ ,
\end{equation}
a short calculation reveals that it differs from 
the standard projection of~$A$.\footnote{
In a previous version of the paper we erroneously identified $A^p$ with
$(1{-}P)A(1{-}P)$. We thank Diego Correa and Fidel Schaposnik for questions
which exposed this confusion to us.} 
To avoid confusion, we call $A^p_\mu$ the `projected' gauge potential.
As the question mark in the above diagram indicates,
it is given only {\it implicitly\/} as a function of~$A_\mu$ through
\begin{equation}\label{AAP}
(1{-}P) \bigl( \pa_{[\mu} A_{\nu]} + [A_\mu\,,A_\nu] \bigr) (1{-}P)\ =\
(1{-}P) (\pa^{\phantom{p}}_{[\mu} A^p_{\nu]} + [A^p_\mu\,,A^p_\nu]+ 
\pa_{[\mu}P\,\pa_{\nu]}P)(1{-}P) \ .
\end{equation}
Similarly to the question of deriving some gauge potential~$A$ from a given
field strength~$F$ in ordinary Yang-Mills theory, it seems difficult to assert 
the existence of~$A^p$ solving~(\ref{AAP}) for a given~$A$. 
Our analogue of the Bianchi identity are the necessary conditions~\footnote{
O.~L. thanks Martin Ro\v cek for a discussion on this point.}
\begin{equation}
(1{-}P)\,\pa^{\phantom{p}}_{[\mu}F^p_{\nu\rho]} P\ =\ 
-F^p_{[\mu\nu} \pa^{\phantom{p}}_{\rho]} P
\qquad\textrm{and}\qquad
(1{-}P)\,\pa^{\phantom{p}}_{[\mu}F^p_{\nu\rho]} (1{-}P)\ =\
\bigl[ F^p_{[\mu\nu}\,,A^p_{\rho]} \bigr] \ ,
\end{equation}
of which the first one is solved by~(\ref{FP1}) 
but the second one is nontrivial.
Nevertheless, from the existence of the canonical solution $A^p{=}A$
for $\th{=}0$ and the smoothness of all expressions in the deformation
parameter~$\th$ we derive some confidence that $A^p$ should exist at least
for sufficiently small values of~$\th$.

\noindent
{\bf Murray-von Neumann transformations.}
After projecting our gauge field $F\mapsto F^p$ via~(\ref{FP1})
and finding a solution $A^p$ of~(\ref{AAP}) we obtain a self-dual
configuration $(A^p,F^p)$ which may be termed the noncommutative 
't~Hooft multi-instanton $(A^p,F^p)$ {\it in a singular gauge\/}.
As is well known in the commutative situation 
(see e.g.~\cite{Rajaraman:1982is}), the singularity in the gauge potential 
can be removed by a singular gauge transformation, 
leading for instance to the BPST~form of the one-instanton solution.
The noncommutative analogues of such singular gauge transformations
are so-called Murray-von Neumann (MvN) transformations.
Indeed, such transformations were proposed in a different context 
-- the modified ADHM approach --
to remove the singularity of the noncommutative one-instanton solution for 
the $U(1)$~\cite{Ho:2000ea} and $U(2)$~\cite{Furuuchi:2001vx} gauge groups.
Here we will show that also for the noncommutative {\it 't~Hooft solutions\/} 
there exist MvN transformations which, in combination with suitable 
projections, repair the deficiency in~(\ref{x}) 
and produce {\it regular\/} solutions for any finite~$n$. 

Given our finite-rank projector~$P$,
we consider a special kind of MvN transformations 
(partial isometry~\cite{Harvey}),
\begin{equation}\label{gauge}
A^p_\mu\ \mapsto\  A'_\mu\ =\ V^\+ A^p_\mu V + V^\+\pa_\mu V
\qquad \textrm{and}\qquad  
F^p_{\mu\nu}\ \mapsto\ F'_{\mu\nu}\ =\ V^\+ F_{\mu\nu}^p V\ ,
\end{equation}
where the intertwining operators 
\begin{equation}
V: \quad \Hcal\otimes\C^2\ \to\ (1{-}P)\Hcal\otimes\C^2
\qquad\textrm{and}\qquad
V^\+: \quad (1{-}P)\Hcal\otimes\C^2\ \to\ \Hcal\otimes\C^2
\end{equation}
satisfy the relation
\begin{equation}\label{V}
V^\+\,V\ =\ 1 \qquad\textrm{while}\qquad
V\,V^\+\ =\ 1-P
\end{equation}
upon extension to $\Hcal\otimes\C^2$.
The configuration~$(A^p,F^p)$ lives on the projective Fock module 
$(1{-}P)\Hcal\otimes\C^2$, but $(A',F')$ exists on the free Fock module 
$\Hcal\otimes\C^2$. 
Therefore, the gauge fields are related to their potentials by~(\ref{FP2}) and
\begin{equation}
F'_{\mu\nu}\ =\ \pa^{\phantom{'}}_{[\mu} A'_{\nu]} + [A'_\mu\,,A'_\nu]\ .
\end{equation}

Via the natural embedding of the reduced Fock space into the full Fock space,
we may extend $V$ and $V^\+$ to endomorphisms of $\Hcal\otimes\C^2$ by
declaring $V^\+P=0=VP$.
Although the extended~$V$ is not unitary, the transformations~(\ref{gauge})
can be regarded as a noncommutative version of singular gauge transformations.
Furthermore, we may actually bypass the projection as far as the field strength
is concerned and directly compute
\begin{equation}\label{ext}
F'_{\mu\nu}\ =\ V^\+ F_{\mu\nu} V\ .
\end{equation}

In the following subsections we will demonstrate that 
to the singular configuration~(\ref{Amu}) with $\phi$ from~(\ref{phin}) 
one should apply firstly a projection onto the reduced Fock space
$(1{-}P_{n-1})\Hcal\otimes\C^2$ to obtain the noncommutative 't Hooft
multi-instanton {\it in a singular gauge\/}, 
and secondly an MvN transformation to arrive at the {\it nonsingular\/} 
noncommutative 't Hooft $n$-instanton solution. In short,
\begin{equation}
(A_\mu,F_{\mu\nu})\quad\buildrel{1-P_{n-1}}\over\longmapsto\quad 
(A^p_\mu,F^p_{\mu\nu})\qquad\buildrel{V_n}\over\longmapsto\qquad 
(A'_\mu,F'_{\mu\nu})
\end{equation}
will bring us to a satisfactory instanton configuration.
However, we will not be able to present explicit expressions for
$A^p$ or $A'$.

\noindent
{\bf One-instanton solution.}
Let us consider first the one-instanton
configuration (\ref{phi1}) in (\ref{Amu}), project it onto 
$(1{-}P_0)\Hcal\otimes\C^2$, and construct its
MvN transformation matrix~$V{\equiv}V_1$.
Due to the relation $(1{-}P_0)X_1(1{-}P_0)=0$, the projected field strength
\begin{equation}
F^p_{\mu\nu}\ =\ (1{-}P_0)\,F_{\mu\nu}\,(1{-}P_0)
\end{equation}
is obviously self-dual, but it is singular because ill-defined 
on~$|0,0\>\otimes\C^2$.  As already mentioned, 
we cannot rigorously prove the existence of the `projected' 
gauge potential~$A^p$ via~(\ref{AAP}) but we shall assume it henceforth.

As will be justified below,
the subsequent MvN transformation applied to $(A^p,F^p)$ is conveniently
factorized as $V_1=\check S_1\,U_1$, with
\begin{equation}
\check S_1\ =\ \begin{pmatrix}1&0\\0&S_1\end{pmatrix} 
\qquad\textrm{and}\qquad
U_1\ =\ \begin{pmatrix} \bar z_\eps & \bar y \\ y & -z_\eps \end{pmatrix}
\frac{\i}{r}\ ,
\end{equation}
where
\begin{equation} \label{shift}
S_1^\+\,S_1\ =\ 1 \qquad\textrm{while}\qquad
S_1\,S_1^\+\ =\ 1-P_0 \ .
\end{equation}
These relations are satisfied by the shift operator
\begin{align} \label{s1}
S_1\ &=\ \sum_{k\ge0} |k{+}1\>\<k| \ ,
\end{align}
constructed from the integer ordered states~(\ref{fock}).
Another possible realization of $S_1$ is~\cite{Furuuchi:2000vc}
\begin{align} \label{s2}
S_1\ &=\ 1 +\sum_{n_2\ge0} \Bigl( |0,n_2{+}1\>\<0,n_2|-|0,n_2\>\<0,n_2| \Bigr)
\ .
\end{align}
A third possible choice (which we shall use) is a `mixture' of (\ref{s1})
and~(\ref{s2}). Namely, recall that the index~$k$ in~(\ref{fock}) introduces 
an integer ordering of states in the two-oscillator Fock space~$\Hcal$:
\begin{equation} \label{ordering}
\bigl\{ |k\> \bigr\}\ =\ \bigl\{ |0,0\>,|0,1\>,|1,0\>,|0,2\>,\ldots\bigr\}\ .
\end{equation}
For fixed instanton number~$n{>}1$ let us move the states~$|0,i\>$ with
$0<i\le n{-}1$ in the sequence~(\ref{ordering}) to the left and enumerate
this new order by~$\hat{k}$,
\begin{equation} \label{neworder}
\bigl\{ |\hat{k}\> \bigr\}\ =\ 
\bigl\{ |0,0\>,|0,1\>,\ldots,|0,n{-}1\>,|1,0\>,\ldots\bigr\}\ .
\end{equation}
Then, the shift operator
\begin{align}\label{s3}
S_1\ &=\ \sum_{k\ge0} |\hat{k}{+}1\>\<\hat{k}|
\end{align}
with the new ordering of states will also satisfy~(\ref{shift}).

We will now confirm our claim that 
\begin{equation} \label{A'}
A'_\mu\ =\ V^\+_1 A^p_\mu V_1 + V^\+_1\pa_\mu V_1 
\qquad\textrm{and}\qquad
F'_{\mu\nu}\ =\ V^\+_1 F_{\mu\nu}^p V_1\ \equiv\ V^\+_1 F_{\mu\nu} V
\end{equation}
brings us to the nonsingular noncommutative one-instanton solution.\footnote{
It is interesting to note that 
$\check S_1^\+ A^p \check S_1+\check S_1^\+ \diff\check S_1$
corresponds to the (singular) $U(2)$ one-instanton configuration produced 
by the modified ADHM construction~\cite{Furuuchi:2001vx}.
It still fails to be self-dual on~$P_0\Hcal$.}
With the help of the definitions (cf. \cite{Furuuchi:2001vx})
$$
r_0\ :=\ \bigl( r^2 -2\th \bigr)^\frac{1}{2}\ =\
\bigl(\bar yy + \bar z_\eps z_\eps \bigr)^\frac{1}{2}\qquad\textrm{and} 
$$
\begin{equation}
r_0^{-1}\ :=\ 
(1{-}P_0)\bigl( \bar yy + \bar z_\eps z_\eps \bigr)^{-\frac12}(1{-}P_0)\ =\
\frac{1}{\sqrt{2\th}}\sum_{n_1,n_2\ne 0}
\frac{|n_1n_2\>\<n_1n_2|}{\sqrt{n_1{+}n_2}}
\end{equation}
it is readily checked that
\begin{equation} \label{nonunitary}
V^\+_1\,V_1\ =\ 1\otimes {\mathbf 1}_2 \qquad\textrm{but}\qquad
V_1\,V^\+_1\ =\ \begin{pmatrix}1{-}P_0&0\\0&1{-}P_0\end{pmatrix}\ =\
(1-P_0)\otimes {\mathbf 1}_2\ .
\end{equation}
{}Furthermore, one finds that
\begin{equation}
X_1\,V_1\ =\ -\frac{\Lambda^2}{2\th(\Lambda^2+2\th )}P_0\,V_1\ =\ 0\ ,
\end{equation}
which assures that
the anti-self-dual part of~$F'_{\mu\nu}$, being proportional to 
$\bar\eta^a_{\mu\nu}V_1^\+\sigma_a X_1 V_1$,
vanishes. We emphasize that the projection $F\mapsto F^p$ is 
not used explicitly in this computation but nevertheless required 
if we want $F'$ to come from a gauge potential~$A'$ which derives 
from~$A^p$ by an MvN transformation.

Obviously, the MvN transformation~(\ref{A'}) extends
the self-duality of~$F_{\mu\nu}^p$ to the whole Fock space 
$\Hcal\otimes\C^2$ . 
This is exactly what was desired to cure the incompleteness 
of the configuration~(\ref{Amu}) with $\phi$ from~(\ref{phi1}).
In combined form, our complete one-instanton gauge field reads
\begin{align}
F'_{\mu\nu}\ &=\
\frac{1}{r}\,
\Bigl(\begin{smallmatrix} z_\eps &  \bar y S_1^\+      \\[4pt]
                          y      & -\bar z_\eps S_1^\+ \end{smallmatrix}\Bigr)
\Bigl(\pa_\mu A_\nu (\phi_1)- \pa_\nu A_\mu (\phi_1) +
      [A_\mu (\phi_1), A_\nu (\phi_1)] \Bigr)
\Bigl(\begin{smallmatrix} \bar z_\eps & \bar y       \\[6pt]
                          S_1 y       & -S_1 z_\eps  \end{smallmatrix}\Bigr)
\,\frac{1}{r}
\end{align}
with abbreviations from (\ref{phi1}), (\ref{ncr}), and (\ref{s1}).
Here, $A_\mu (\phi_1)$ is obtained by substituting 
(\ref{phi1}) into~(\ref{Amu}).

In the commutative limit, 
our configuration $(A',F')$ clearly reduces to the BPST~instanton 
because $\check S_1=\textrm{diag}(1,S_1)$ becomes a unitary matrix 
and the remaining part of the MvN transformation (mediated by $U_1$) 
turns {\it exactly\/} into the singular gauge transformation 
from the 't~Hooft to the BPST gauge (see~\cite{Rajaraman:1982is}). 
For illustration, the operator
$S_1$ from~(\ref{s2}) can be rewritten as (cf.~\cite{Ho:2000ea})
\begin{equation}
S_1\ =\ \bar y\,(\bar y y +2\th)^{-1}y\ +\
(1 -\bar y\,(\bar y y +2\th)^{-1}y)\,\bar z_\eps\, 
(\bar z_\eps z_\eps + 2\th)^{-\frac12}\ ,
\end{equation}
from which one easily sees that in the commutative limit
it approaches the identity.

\noindent
{\bf Multi-instanton solutions.}
Again, a projection (by $1{-}P_{n-1}$) plus a Murray-von Neumann 
transformation (by $V_n$) will enable us to extend the projected self-dual 
solution to the complete Fock space. 
We decompose $V_n=\check S_n\,U_n$ and propose
\begin{equation}\label{gauge5}
\check S_n\ =\ \begin{pmatrix}1&0\\0&S_n\end{pmatrix} 
\qquad\textrm{and}\qquad
U_n\ =\ \U\,U_1^n\,{\U}^\+ \ ,
\end{equation}
where
\begin{equation}
\U\ :=\ \sum_{i\ge0} |h_{i+1}\>\<\hat{i}|
\end{equation}
is the unitary transformation from the basis $\{|\hat{i}\>\}$ to the 
basis~$\{|h_{i+1}\>\}$ in~$\Hcal$, 
and $S_n: \Hcal\to(1{-}P_{n-1})\Hcal$ is to satisfy the relations
\begin{equation} \label{Sn}
S^\+_{n}\,S_{n}\ =\ 1 \qquad\textrm{and}\qquad
S_{n}\,S^\+_{n}\ =\ 1-P_{n-1}\ =\ 1-\sum_{i=1}^n |h_i\>\<h_i|\ .
\end{equation}
In the orthonormal basis one can choose $S_n$ and $S^\+_n$ in the form
\begin{equation}
S_{n}\ =\ \sum_{i\ge 1}\,|h_{i+n}\>\<h_i|\qquad\textrm{and}\qquad
S^\+_{n}\ =\ \sum_{i\ge 1}\,|h_{i}\>\<h_{i+n}|\ .
\end{equation}
Since this basis was constructed from the states $|b_1\>,\ldots,|b_n\>$,
the operator $S_n$ contains all the information about the position parameters 
$b_1,\ldots,b_n$.
It is easy to see that
\begin{equation}
S_n\ =\ \U\,S_1^n\,{\U}^\+
\end{equation}
with $S_1$ from~(\ref{s3}).

Let us justify our proposal~(\ref{gauge5}).
Employing the relations
\begin{equation}
U_1^n\ =\ 1\ - \sum_{i=0}^{n-1} \begin{pmatrix}
|\hat{i}\>\<\hat{i}| & 0 \\ 0 & 0 \end{pmatrix}
\end{equation}
with
\begin{equation}
\U\Bigl( \sum_{i=0}^{n-1} |\hat{i}\>\<\hat{i}| \Bigr) {\U}^\+ \ =\
\sum_{i=0}^{n-1} |h_{i+1}\>\<h_{i+1}|\ =\ P_{n-1}
\end{equation}
one can show that
\begin{equation}
U_n^\+\,U_n\ =\ 1\otimes {\mathbf 1}_2 \qquad\textrm{but}\qquad
U_n\,U_n^\+\ =\
\begin{pmatrix}1{-}P_{n-1}&0\\0&1\end{pmatrix}\ .
\end{equation}
Together with~(\ref{Sn}) one gets
\begin{equation}
V_n^\+\,V_n\ =\ 1\otimes {\mathbf 1}_2 \qquad\textrm{but}\qquad
V_n\,V_n^\+\ =\ 
\begin{pmatrix}1{-}P_{n-1}&0\\0&1{-}P_{n-1}\end{pmatrix}
\ =\ (1-P_{n-1})\otimes {\mathbf 1}_2\ .
\end{equation}
{}Finally, substituting (\ref{phin}) into (\ref{Amu}),
computing its field strength~$F_{\mu\nu}$ and performing 
the (extended) MvN transformation~(\ref{ext}) by the matrices~(\ref{gauge5}) 
we learn that the anti-self-dual part of~$F'_{\mu\nu}$ 
indeed vanishes everywhere in $\Hcal\otimes\C^2$, again due to
$X_nV_n=0=V^\+_nX_n$. Hence, our final result
\begin{equation}
A'_\mu\ =\ V^\+_n\,A^p_\mu\,V_n + V^\+_n\,\pa_\mu\,V_n
\qquad\textrm{and}\qquad
F'_{\mu\nu}\ =\ V^\+_n\,F_{\mu\nu}^p\,V_n\ \equiv\ V^\+_n\,F_{\mu\nu}\,V_n
\end{equation}
constitutes a proper noncommutative generalization 
of the 't~Hooft $n$-instanton solution. 
It has topological charge $Q=n$ since in the $\th\to0$ limit our 
solution coincides with the standard 't~Hooft solution, and the 
topological charge does not depend on $\th$. This may also be shown by 
reducing the action integral to the trace of the projector $P_{n-1}$, 
following Furuuchi~\cite{Furuuchi:2000vc}.

\section{Concluding remarks}

\noindent
Proper noncommutative instantons are constructed by not only replacing 
the coordinates in the commutative configuration with their operator analogues 
but also applying a projection and an appropriate MvN transformation.
We have demonstrated this beyond the (previously considered) case of the 
one-instanton solution, providing explicit formulae for the field strength
(but not the gauge potential) of regular noncommutative 't~Hooft 
multi-instantons in~$U(2)$.  Our results are easily generalized to~$U(N)$.
We have pointed out that the MvN transformation is needed to
remove the source singularities in the reduced SDYM equation,
which hamper self-duality on some subspace. 
In order to take advantage of the MvN transformation, however, we had to 
pass from the original gauge potential to a `projected' gauge potential 
in an implicit manner, thereby foregoing an explicit solution for it.

We found it easy to work with the twistor approach
because it and the Atiyah-Ward ans\"atze directly generalize 
to the noncommutative realm, providing us with a systematic
and straightforward strategy for the construction of
self-dual gauge-field configurations. In this method, the main task is to find 
two holomorphic (in~$\l$ and $\tilde\l$, respectively)
regular matrix-valued operators $\psi_+$ and $\psi_-$ such that 
their `ratio'~$\psi^{-1}_{+}\psi_{-}$
defines a holomorphic bundle over noncommutative twistor space 
with appropriate global properties.
In fact, finding solutions to the splitting problem is not
made any harder by noncommutativity.

Multi-instantons can also be obtained by employing the dressing
approach~\cite{Belavin:1978pa, Forgacs:1981su}. 
In its noncommutative variant,
one is to find a meromorphic (in~$\l$) matrix-valued operator~$\psi$
(having finite-order poles in the spectral parameter~$\l$) which
obeys some linear differential equations.
The noncommutative dressing method was successfully applied
to the study of noncommutative solitons in a $2{+}1$ dimensional
integrable field theory~\cite{Lechtenfeld:2001uq,Lechtenfeld:2001aw}.
It would be illuminating to also exercise it on the subject of
noncommutative instantons and to compare the results
with those obtained in the twistor approach.

{}Finally, noncommutative instantons are interpreted as
D$p$-branes within coincident D$(p{+}4)$-branes carrying a
constant two-form $B$-field background~\cite{Seiberg:1999vs}. 
Thus, they have immediate bearing on the issue of 
nonperturbative string backgrounds.

\vfill\eject
%\bigskip
\noindent
{\large{\bf Acknowledgements}}

\smallskip
\noindent
We are grateful to D. Correa and F. Schaposnik for useful comments on
an earlier version of this paper.
O.~L. thanks M. Ro\v cek for fruitful discussions 
and reading of the manuscript.
This work is partially supported by DFG grant Le~838/7-1
and by a sabbatical research grant of the Volkswagen-Stiftung.

\bigskip

\setcounter{section}{0}
\renewcommand{\thesection}{\Alph{section}}
\section{Noncommutative 't~Hooft instantons from the ADHM approach}

\noindent
{\bf ADHM construction.}
Here we restrict ourselves to the case of self-dual noncommutative
Euclidean space~$\R^4_\th$ with the tensor~(\ref{asdth}) given by
\begin{equation}
\th^{\mu\nu}\ =\ \th\,\h^{3\mu\nu}
\end{equation}
where
\begin{equation}
\eta^a_{\mu\nu}\ =\ \begin{cases}
\eps^a_{bc} & \textrm{for} \quad \mu =b\,,\ \nu =c \\
\de^a_\mu  & \textrm{for} \quad \nu =4 \\
-\de^a_\nu   & \textrm{for} \quad \mu =4 \end{cases}
\end{equation}
is the self-dual 't~Hooft tensor.

Let us introduce the matrices
\begin{equation}
\bigl(e_\mu\bigr)\ =\ \bigl(-\i\s_a\,,\,1\bigr)
\qquad\textrm{and}\qquad
\bigl(e^\+_\mu\bigr)\ =\ \bigl(\i\s_a\,,\,1\bigr)
\end{equation}
which enjoy the properties
\begin{align}
e^\+_\mu\,e_\nu\ &=\ \de_{\mu\nu} + \h^a_{\mu\nu}\,\i\s_a\
=:\ \de_{\mu\nu} + \h_{\mu\nu} \ ,\\ 
e_\mu\,e^\+_\nu\ &=\ \de_{\mu\nu} + \bar{\h}^a_{\mu\nu}\,\i\s_a\
=:\ \de_{\mu\nu} + \bar{\h}_{\mu\nu} \ .
\end{align}
Using these matrices one can introduce $x:=x^\mu e^\+_\mu$
with $\{x^\mu\}\in\R^4_\th$.

The (modified) ADHM construction (see 
\cite{Nekrasov:1998ss,Ho:2000ea,Furuuchi:2001vx,Nekrasov:2000ih,Chu:2001cx})
of an $n$-instanton solution is based on a $(2n{+}2)\times2$ matrix~$\Psi$
and a $(2n{+}2)\times2n$ matrix $\Delta=a+b(x{\otimes}{\bf1}_n)$
where $a$ and~$b$ are constant $(2n{+}2)\times2n$ matrices.
These matrices must satisfy the following conditions:
\begin{align}
\Delta^\+\Delta\quad& \textrm{is invertible}\ ,\label{c1}\\
[\,\Delta^\+\Delta\,,\,e_\mu\otimes{\bf1}_n\,]\ &=\ 0 \quad\forall x\ , 
\label{c2}\\
\Delta^\+\Psi\ &=\ 0\ ,\label{c3}\\
\Psi^\+\Psi\ &=\ {\bf1}_2 \ .\label{c4}
\end{align}
It is not difficult to see that conditions (\ref{c1}) and~(\ref{c2})
are met if
\begin{equation} \label{c5}
\Delta^\+\Delta\ =\ {\bf1}_2 \otimes f^{-1}_{n\times n}\ .
\end{equation}
For $(\Delta,\Psi)$ satisfying~(\ref{c1})--(\ref{c4}) 
the gauge potential is chosen in the form
\begin{equation} \label{adhmA}
A\ =\ \Psi^\+\,d\Psi\ .
\end{equation}
The resulting gauge field~$F$ will be self-dual if $\Delta$ and~$\Psi$
obey the completeness relation
\begin{equation} \label{complete}
\Psi\,\Psi^\+\ +\ \Delta\,(\Delta^\+\Delta)^{-1}\Delta^\+\ =\ {\bf1}_{2n+2}\ .
\end{equation}

\noindent
{\bf Ansatz.}
For constructing noncommutative 't~Hooft $n$-instantons let us take
(cf.~\cite{Corrigan:1978ce,Osborn:1978rn})
\begin{equation}
\Psi\ =\ \begin{pmatrix}
\Psi_0 \\ \Psi_1 \\ \vdots \\ \Psi_n \end{pmatrix}
\ ,\qquad
a\ =\ \begin{pmatrix}
\La_1{\bf1}_2 & \ldots & \La_n{\bf1}_2 \\
-b_1          &        & {\bf0}_2      \\
              & \ddots &               \\
{\bf0}_2      &        & -b_n          \end{pmatrix}
\qquad\textrm{and}\qquad
b\ =\ \begin{pmatrix}
{\bf0}_2 & \ldots & {\bf0}_2 \\
{\bf1}_2 &        & {\bf0}_2 \\
         & \ddots &          \\
{\bf0}_2 &        & {\bf1}_2 \end{pmatrix}
\end{equation}
with $b_i=b_i^\mu e^\+_\mu$. It follows that
\begin{equation} \label{c5res}
\Delta^\+\Delta\ =\ {\bf1}_2 \otimes (\de_{ij}r_j^2 + \La_i\La_j)\
=:\ {\bf1}_2 \otimes (R + \La\La^T)\ ,
\end{equation}
where
\begin{equation}
R\ =\ \begin{pmatrix}
r_1^2 &        & 0     \\
      & \ddots &       \\
0     &        & r_n^2 \end{pmatrix}\ ,\qquad
\La\ =\ (\La_1,\ldots,\La_n)\ ,\qquad
r_j^2\ =\ \de_{\mu\nu}\,(x^\mu{-}b_j^\mu)(x^\nu{-}b_j^\nu)\ ,
\end{equation}
and the $\La_i$ are constants parametrizing the scale of the $i$th instanton.
{}From~(\ref{c5res}) we see that the condition~(\ref{c5})
(and thus also (\ref{c1}) and~(\ref{c2})) is satisfied.
Indeed, by direct calculation one finds
\begin{equation}
{\bf1}_2\otimes f_{n\times n}\ =\ (\Delta^\+\Delta)^{-1}\ =\
{\bf1}_2\otimes\bigl( R^{-1} - R^{-1}\La\,\phi_n^{-1}\La^T R^{-1} \bigr)
\end{equation}
with
\begin{equation}
\phi_n\ =\ 1+\ \sum_{i=1}^n \frac{\La_i^2}{r_i^2}\ .
\end{equation}

For the given $\Delta=a+b(x{\otimes}{\bf1}_n)$ 
and $x_i:=(x^\mu{-}b_i^\mu)e^\+_\mu$, the condition~(\ref{c3}) becomes
\begin{equation}
\La_i\Psi_0 + x_i^\+\Psi_i\ =\ {\bf0}_2 \qquad\textrm{for}\quad i=1,\ldots,n\ .
\end{equation}
These equations are solved by
\begin{equation} \label{Psisol}
\Psi_0\ =\ \phi_n^{-\frac12}\,{\bf1}_2 \qquad\textrm{and}\qquad
\Psi_i\ =\ -x_i\,\frac{\La_i}{r_i^2}\,\phi_n^{-\frac12}
\end{equation}
where the factor $\phi_n^{-\frac12}$ was introduced 
to achieve the normalization
\begin{equation}
\Psi^\+\Psi\ =\ \phi_n^{-\frac12}\Bigl(
1+\ \sum_{i=1}^n\frac{\La_i^2}{r_i^2} \Bigr){\bf1}_2\;\phi_n^{-\frac12}\ 
=\ {\bf1}_2\ .
\end{equation}
Hence, our $(\Delta,\Psi)$ satisfies all conditions (\ref{c1})--(\ref{c5}),
and we can present the gauge potential~(\ref{adhmA}).

\vfill\eject

\noindent
{\bf Completeness relation.}
Being convinced that they have constructed self-dual field configurations,
many authors stop at this point 
and ignore the completeness relation~(\ref{complete}).
However, the latter may be violated, in which case one obtains a 
{\it singular\/} field configuration (cf.~the discussion in~\cite{Chu:2001cx}).
Indeed, our example solution~(\ref{Psisol}) indicates just that.
Namely, after (lengthy) computations we arrive at
\begin{equation} \label{incomplete}
\Psi\,\Psi^\+\ +\ \Delta\,(\Delta^\+\Delta)^{-1}\Delta^\+\ 
=\ {\bf1}_{2n+2} - \Pi\ ,
\end{equation}
where $\Pi$ is a projector,
\begin{equation}
\Pi\ :=\ \begin{pmatrix}
{\bf0}_2 &         &        & {\bf0}_2 \\
         & {\bf1}_2-x_1\sfrac{1}{r_1^2}x_1^\+ &        &        \\
         &         & \ddots &          \\
{\bf0}_2 &         &        & {\bf1}_2-x_n\sfrac{1}{r_n^2}x_n^\+
\end{pmatrix}
\quad,\quad\textrm{and}\qquad
{\bf1}_2-x_i\frac{1}{r_i^2}x_i^\+\ =\ \begin{pmatrix}
|b_i\>\<b_i| & 0 \\ 0 & 0 \end{pmatrix}\ .
\end{equation}
Here, the $|b_i\>$ are the coherent states defined in~(\ref{bstates}).
Notice that in the commutative limit $\Pi\to0$, and the completeness
relation will be saturated.

\noindent
{\bf Field strength.}
We finally evaluate the field strength. Substituting~(\ref{Psisol})
into~(\ref{adhmA}) and using~(\ref{incomplete}) we find 
\begin{align}
F_{\mu\nu}\ &=\
\pa_\mu(\Psi^\+\pa_\nu\Psi)\ -\ \pa_\nu(\Psi^\+\pa_\mu\Psi)\
+\ [\,\Psi^\+\pa_\mu\Psi\,,\,\Psi^\+\pa_\nu\Psi\,] 
\nonumber\\
&=\ (\pa_\mu\Psi^\+) (1-\Psi\Psi^\+)\,\pa_\nu\Psi\ -\ (\mu\leftrightarrow\nu) 
\nonumber\\
&=\ (\pa_\mu\Psi)^\+ 
\bigl( \Delta\,(\Delta^\+\Delta)^{-1}\Delta^\+ + \Pi \bigr)\,\pa_\nu\Psi\ 
-\ (\mu\leftrightarrow\nu) 
\nonumber\\
&=\ \Psi^\+(\pa_\mu\Delta)\,(\Delta^\+\Delta)^{-1} (\pa_\nu\Delta^\+)\Psi\
+\ (\pa_\mu\Psi)^\+\,\Pi\;\pa_\nu\Psi\ -\ (\mu\leftrightarrow\nu) 
\nonumber\\
&=\ \Psi^\+\,b\,e^\+_\mu (\Delta^\+\Delta)^{-1} e_\nu\,b^\+\Psi\ 
+\ \phi_n^{-\frac12} e_\mu
\Bigl(0,-\sfrac{\La_1}{r_1^2},\ldots,-\sfrac{\La_n}{r_n^2}\Bigr)\,\Pi
\begin{pmatrix}
0 \\ -\sfrac{\La_1}{r_1^2} \\ \vdots \\ -\sfrac{\La_n}{r_n^2}
\end{pmatrix}
e^\+_\nu\,\phi_n^{-\frac12}\ -\ (\mu\leftrightarrow\nu) 
\nonumber\\
&=\ 2\,\Psi^\+ b\,(\Delta^\+\Delta)^{-1} \h_{\mu\nu}\,b^\+\Psi\ +\
\phi_n^{-\frac12}\biggl(\sum_{i=1}^n\frac{\La_i^2}{4\th^2}\,|b_i\>\<b_i|\biggr)
\phi_n^{-\frac12}\,2\,\bar\h_{\mu\nu} 
\nonumber\\
&=\ 2\,\Psi^\+ b\,(\Delta^\+\Delta)^{-1} \h_{\mu\nu}\,b^\+\Psi\ 
-\ X_n\,2\,\bar\h_{\mu\nu}\ ,
\end{align}
where $X_n$ is precisely the source term~(\ref{x}) derived earlier in the
twistor approach. We conclude that the anti-self-dual part of~$F_{\mu\nu}$
is nonzero and, in complex coordinates, coincides with the r.h.s. 
of~(\ref{csdym}). Hence, the ADHM approach encounters exactly the same 
obstacle as the twistor method.

\vfill\eject

}

\end{document}